# Multi-agent systems and decentralized artificial superintelligence

Ponomarev S. (1), Voronkov A. E. (2)

(1) Moscow Power Engineering Institute

(2) Moscow Institute of Physics and Technology

**Abstract**

Multi-agents systems communication is a technology, which provides a way for multiple interacting intelligent agents to communicate with each other and with environment. Multiple-agent systems are used to solve problems that are difficult for solving by individual agent.

Multiple-agent communication technologies can be used for management and organization of computing fog and act as a global, distributed operating system.

In present publication we suggest technology, which combines decentralized P2P BOINC general-purpose computing tasks distribution, multiple-agents communication protocol and smart-contract based rewards, powered by Ethereum blockchain.

Such system can be used as distributed P2P computing power market, protected from any central authority.

Such decentralized market can further be updated to system, which learns the most efficient way for software-hardware combinations usage and optimization. Once system learns to optimize software-hardware efficiency it can be updated to general-purpose distributed intelligence, which acts as combination of single-purpose AI.

**Introduction**

This article outlines the theoretical prerequisites of creating intelligent multi-agent system of Decentralized Artificial Intelligence «SONM», as well as discusses the practical realization of these ideas. If you are interested in the theoretical questions, then see <u>chapter one</u>. If you are only interested in the practical realization, skip to <u>chapter two</u>.

## Chapter 1. The theoretical rationale for the existence of decentralized artificial intelligence

**Agent in Informatics and artificial intelligence**

How is interpreted the term "agent" in modern Informatics and AI? To date, several different interpretations of this term have been formed (Fig. 1.1).

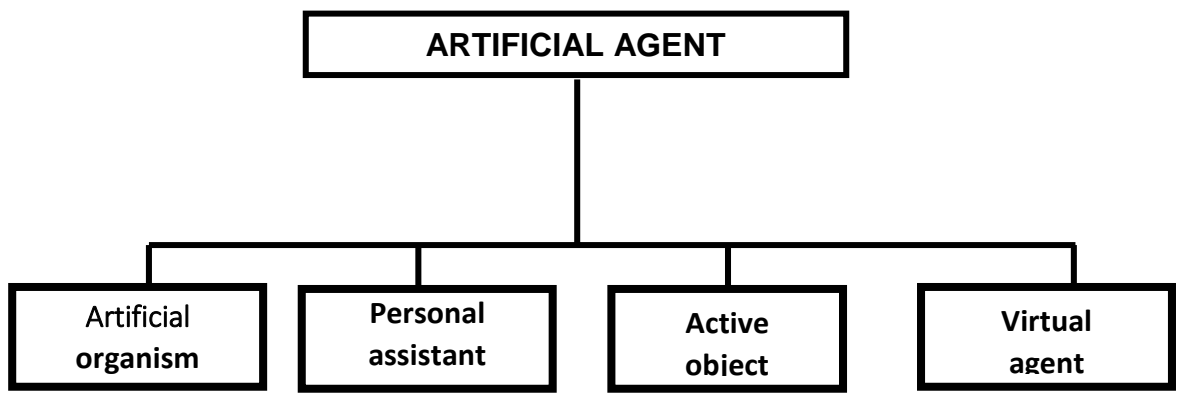

Fig. 1.1. Possible interpretations of the concept of "artificial agent".

Firstly, J. Holland [1] was first who introduced an idea about agent as **an artificial organism**, developing in the population of its kind, tending to learn and adapt to the environment in order to survive in it (and to defeat the competitors). This interpretation of agent is based on the theoretical approaches and models of artificial evolution (mutations and modifications of agents, and their struggle for survival, the selection of the strongest or rejection of the weakest) and the principles of artificial life (self-reproduction, self-preservation, self-determination, self-control of agents, and so on). It is closely related to robotics (the construction of integral robots and functioning of robot groups), the problems of "armor" and "weapon" relations in computer networks, information security and information attacks, computer virology and creating **Liveware** tools (evolving software, which is constructed considering principles and mechanisms of the behavior of living organisms).

Secondly, a metaphor appeared which interpreted agent as **a personal assistant** of user or, later, as **intelligent intermediary** between the user and the environment in which he works [2]. In particular, the strategy of artificial agent development, outlined in the **IBM White Paper**, is based on this idea of a "personal assistant", where the agent is any software or hardware system that can operate in order to achieve the objectives set by the user.

The idea of **personal assistants** in work of users with computers in its simplest form was embodied in a number of popular software products in late 90s. For example, Microsoft has built Wizards and System Agent in Windows95 and paperclip-assistant Clippit has appeared in Microsoft Office. Mac S includes a learning agent Open Sesame!, Lotus Notes V4 also has built-in agents. Among the modern personal assistants a voice helper **Siry**, existing in Apple's products can be distinguished.

Thus, agents are often understood as autonomous (or semi-autonomous) software modules that can collaborate with the user and adapt to him. "**Semi-autonomous**" in this context means a software agent dependence on user, particularly, the user's ability to change the level of autonomy of his agent. This ensures not only friendly, but also **personified** character of the user interface. The origins of this concept of agent relate to the theory of dialogue "human-computer" and means of intelligent interface development.

Real boom in the field of **software agents** began with the development of the Internet and appearance of **chat bots**. Information agents, such as PointCast, deliver news to users and report changes in selected sites. Shopping agents, such Bargain Finder, working for users, compare prices in electronic shops. Robotic spiders (**crawlers**) roam in the links and index an information for the search engines. Robots - chat bots are able conclude a transaction with the users *to exchange cryptocurrency*, etc.

The greatest prospects for further applications of personalized "agents – user assistants" are associated with targeted information search in Internet, given its semantic and pragmatic characteristics, as well as the support of multi-criterial hard-to-formalize decision. We should expect a recently appeared concept "agentware" that characterizes the new architectural principles of information processing based on agent, will become widespread.

Thirdly, the agents may be considered as **active objects** or **meta-objects** endowed with a certain degree of subjectness, i.e. able to manipulate other objects (e.g., **smart contracts**), create and destroy them, and interact with the environment and other agents. In this context, they can be created on the basis of constraint programming using the Active Object technologies [3]. Thus, the software technology of agents and agent-oriented programming are understood in general as a natural development of the ideas of object-oriented programming (OOP). The agent is a "self-contained programmed process, which includes some state and has the opportunity to interact with other agents via **exchanging messages**" [4]. Accordingly, the agent-oriented programming (AOP) [5] is a new programming paradigm based on "social points of view" on the calculations.

**In total** − personalized assistants like **Siry**; programs-"demons" in the systems of UNIX and Windows utilities; softbots - chat-bots in Telegram, crawlers in search engines, programs-advisors in Forex and Metatrader, bot players in computer games, smart contracts in Ethereum, etc. may be provided as the examples of agents.

**Multi-agent systems**

Solving of the task by one agent on the basis of knowledge engineering is a point of view of classical AI. According to it, the agent (for example, the intelligent system), having a global vision of the problem, has all the necessary skills, knowledge and resources for solving the task. In contrast, when creating multi-agent systems (MAS) we assume that a single agent can have only a partial understanding of the task and can solve only some of its subtasks. Therefore, to solve any complex problem, as a rule, the **interaction** of agents is required, which is inseparable from the formation of MAS. The tasks in MAS are distributed between the agents, each of which is considered as a member of the group or organization. Distribution of tasks involves assigning roles to each of the agents, the definition of measure of its responsibility and requirements to its experience.

Depending on whether the distribution comes from the set task of or the ability of the particular agent, can be distinguished the systems of a distributed solution of the tasks and systems of decentralized AI. In the first case, the process of decomposition of the original task and the reverse process of composition of the obtained solutions is centralized. MAS is rigidly projected downward on the basis of partitioning of the general task into separate, relatively independent subtasks and preliminary determination of the agents' roles (or pre-formulated requirements to them). In the second case, the distribution of tasks happens largely spontaneously, directly in process of the interaction of the agents.

Activity of artificial (computer) systems and organization of their joint work related to the collective and concerted solution of the tasks in virtual communities are fundamental characteristics of the conceptual novelty of advanced information technology and network organizations, built on the principles of the MAS.

**The main directions of development of multi-agent systems, distributed artificial intelligence**

Synergistic content of MAS conception is based on the processes of interaction of individual and collective agents, leading to the formation of artificial groups and communities, i.e., **social computing systems with fundamentally new features**. Depending on the number of interacting agents and the inherent characteristics of their interactions, the various directions of development and types of MAS can be distinguished.

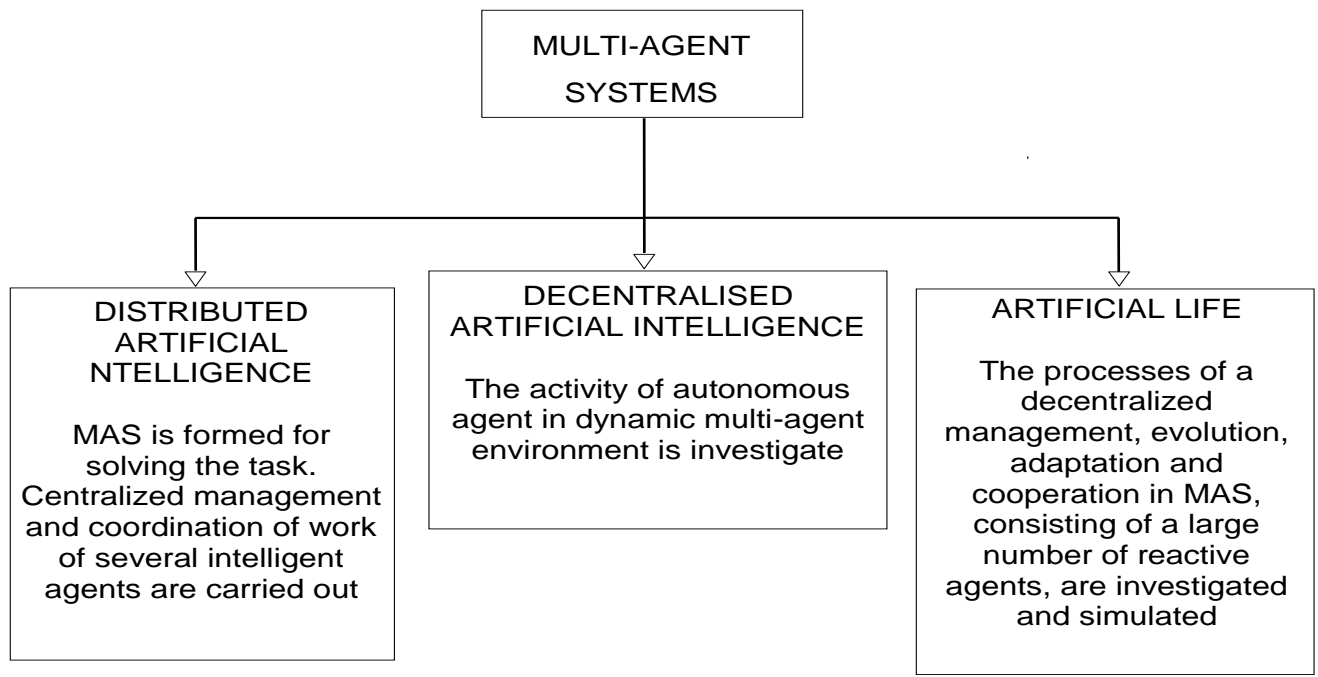

Fig. 1.2. Classification of multi-agent systems.

A **distributed artificial intelligence and artificial life** (in the narrow sense of the term) and the main directions in the development of MAS (Fig. 1.2). The studies of interaction and cooperation of a small number of intelligent agents, for example, the classical intelligent systems, including knowledge bases and solvers, compose a kernel of Distributed Artificial Intelligence (**DAI**) [6−11]. The main problem in **DAI** lies in the development of intellectual groups and organizations, capable to solve tasks by reasoning, which is related to the treatment of symbols. In other words, group intellectual behavior in **DAI** is based on individual intellectual behaviors. This means a congruence of the objectives, interests and strategies of different agents, coordination of their actions, the resolution of conflicts through negotiations; theoretical base in this process consists of the results obtained in the psychology of small groups and the sociology of organizations.

DAI systems are defined by three main characteristics: 1) a method for the distribution of tasks between agents; 2) a method of distribution of powers; 3) method of communication of the agents.

Typical scheme of distributed solution of the tasks by several agents includes the following steps:

1) agent-subordinator (Head, the central body) decomposes the original problem into separate tasks;

2) these tasks are distributed between the agents-executants;

3) each agent-executant solves the task, sometimes also dividing it into sub-tasks;

4) in order to obtain the overall result a composition, integration of particular results corresponding to the selected task is produced. Agent-Integrator is responsible for the overall result (often, this is the same agent-subordinator).

**BOINC as a system of distributed artificial intelligence**

The BOINC (Berkeley Open Infrastructure for Network Computing.) is an open software platform (Berkeley University for GRID computing) − a non-profit middleware for organization of distributed computing. It is used for volunteer computing organization.

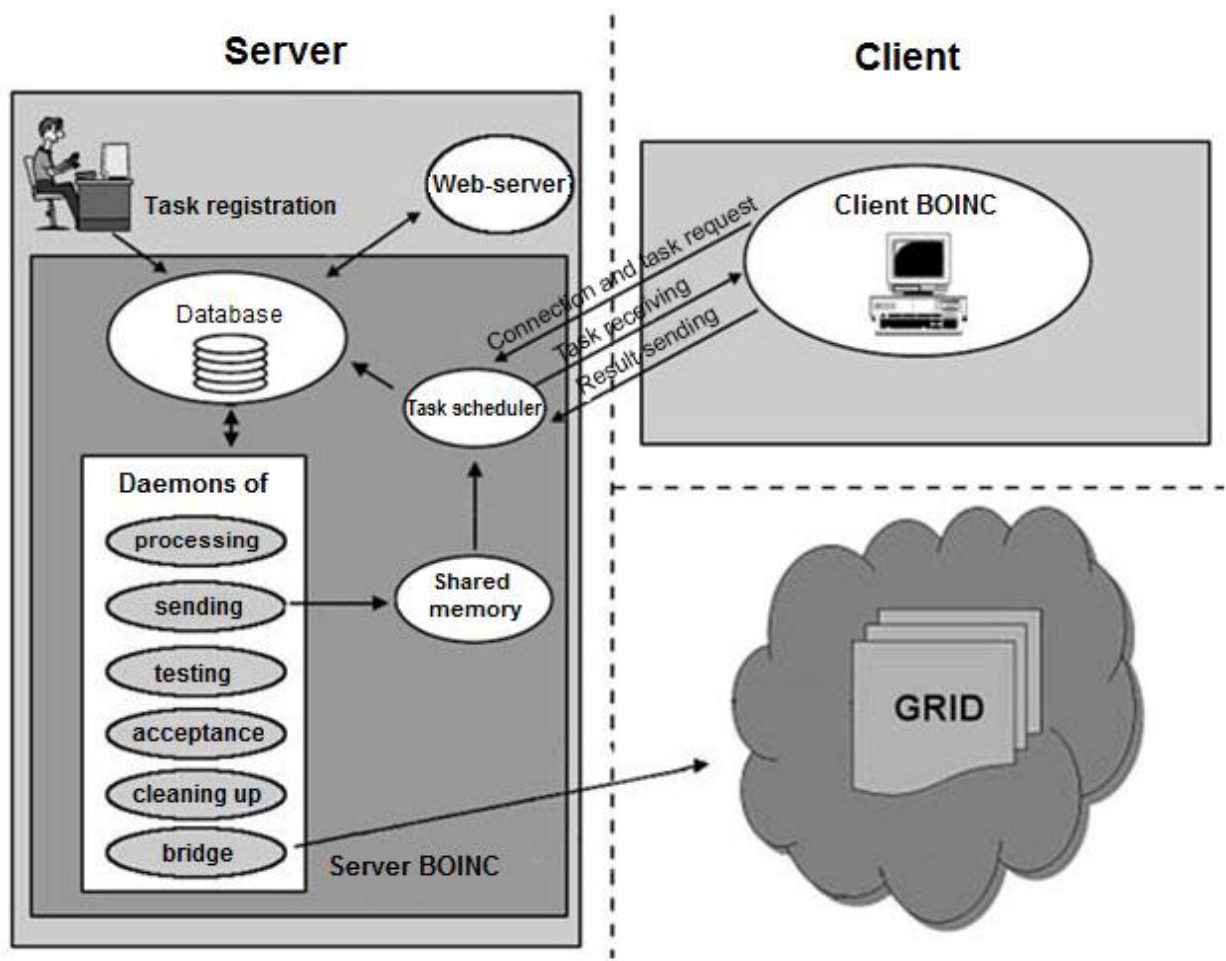

Fig. 1.3. Scheme of BOINC work.

BOINC architecture (Fig. 1.3) is based on the idea of a **distributed AI** - its server consists of a set of individual subsystems (agents), each of which is responsible for its own well-defined task, such as performing of calculations, file transfer, etc. Each subsystem checks the status of sub-tasks produces some actions and changes the state of subtasks --- so they work in an infinite loop.

In general, the system consists of BOINC server, a plurality of clients, performing the tasks of the server and, possibly, additional components in the form of GRID-affiliated networks.

As we can see from the previous section, BOINC system, as well as any other system of distributed AI, existing at the moment, are the centralized systems, strictly managed by a central server, which, of course, is a significant **drawback**.

SONM system uses the BOINC system, which is a system of **distributed** intelligence as the basis for the creation of a **decentralized intelligence.**

**Swarm Intelligence and artificial life**

The second direction − artificial life (AL) – is associated to a greater extent with the interpretation of intellectual behavior in the context of survival, adaptation and self-organization in dynamic, hostile environments, which goes back to the works by Piaget (see [12]).

V.M. Bekhterev noticed that the more elementary goals and objectives of the collective, the more sizes the collective can reach. For example, man in a crowd of people loses inhibition ability, but wins in the imitative ones.

In the tideway of AL a global intelligent behavior of the entire system is considered as a result of local interactions of a large number of simple and not necessarily intelligent agents. Terms such as "**collective intelligence**" (see, e.g., [13]; Fig. 1.4) or "**swarm intelligence**" [14, 15] are also used for AL. Adherents of this direction, in particular, R. Bruks, J. Deneubourg, L. Steels, etc. [16−20] rest on the following provisions: 1) the MAS is a population of simple and mutually dependent agents; 2) each agent independently determines its reaction to the events in the local environment and the interactions with other agents; 3) interrelations between agents are horizontal, i.e., there is no an agent-supervisor, managing the interaction of other agents; 4) there is no precise rules to define the global behavior of agents; 5) the behavior, properties and structure on the collective level are generated by only the local interactions of agents.

Here, mechanisms of reaction to the impact of the environment and local interactions in general case do not include aspects such as forecasting, planning, processing knowledge, but sometimes allow to solve complex problems. Typical biological examples of such collective intelligence include ant colonies, beehives, bird flocks, etc.

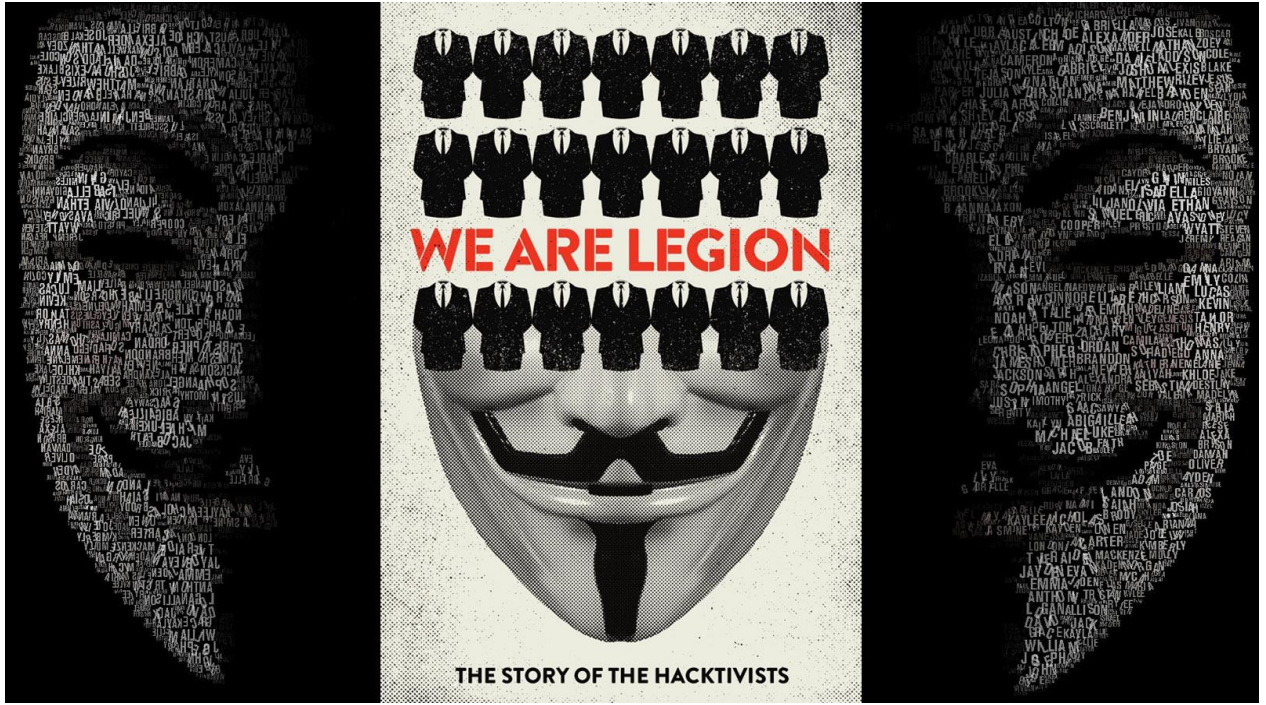

Fig. 1.4. Collective Intelligence.

In program form, blockchain can serve as the most typical example of the existence of swarm intelligence − a lot of miners work as reactive agents, who without any control of agent-supervisor carry out the work on maintenance of the network, moving only in accordance with their own motivation.

**The disadvantages** of such systems consist of the inability of agents to the more complex organization, planning, and solution of the tasks, requiring sequential execution or data analysis, as well as the excess parallelism of tasks' execution.

**Decentralized artificial intelligence as a prologue of the future**

Many authors show **principal** differences between **distributed** and **decentralized AI** (Fig. 1.1) [21]. The ideology of distributed task solving [6, 22, 23] assumes mainly the separation of knowledge and resources between the agents and, to a lesser extent, distribution of management and power; as a rule, it postulates the existence of a common governing body that provides decision-making in critical (conflict) situations. At the same time, an overall complex problem for whose solution a group of agents is formed, a common conceptual model is constructed and the global criteria for achieving the goal are introduced, serves as an original object of the study.

In a fully decentralized systems management takes place only because of local interactions between agents. Here, the basic object of study is not a distributed solution of some general task, but **an activity of autonomous agent in dynamic multi-agent world** (as well as the coordination of different agents). At the same time, local tasks of individual agents, solved on the basis of local conceptual models and local criteria, are described along with the distributed knowledge and resources (Fig. 1.5).

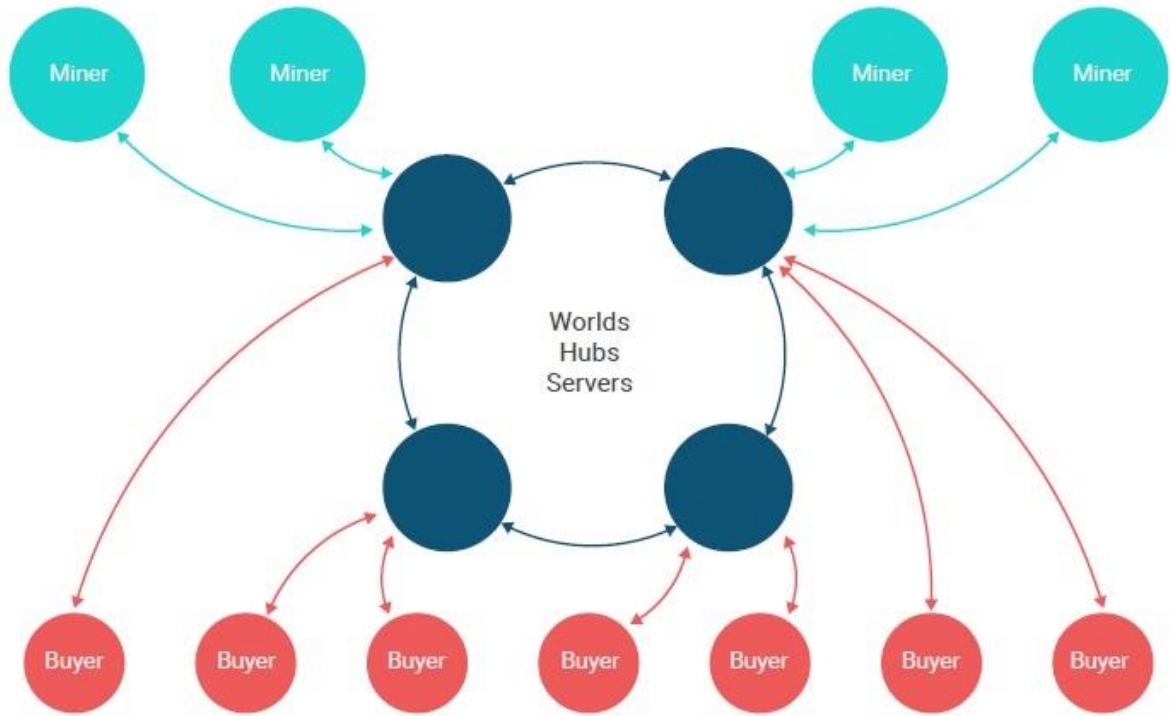

Fig. 1.5. Model of multi-agent system.

*We, Founders of **SONM**, believe that it is time to create such a **decentralized AI** based on existing models of multi-agent systems, such as the **BOINC**, which serves as an example of distributed AI and **blockchain systems**, which serve as examples of "swarm intelligence". Combining these two technologies will allow to compensate for the shortcomings of these systems and to enhance their benefits, thus creating **a completely new** AI model that will work for the benefit of all humankind.*

## *Chapter 2. The practical implementation of the system*

Therefore, in the previous chapter, we found that an agent within the AI will be called a bot (robot, chat bot, softbot, etc.), which is motivated, able to perform actions with the external environment, and most importantly, able to communicate with other bots and shape their behavior.

Multi-agent systems are systems consisting of these agents, which, however, are not always artificial.

In addition, we found that the main directions of development of multi-agent systems are following:

- Swarm intelligence;
- Distributed intelligence;
- Decentralized intelligence.

Swarm intelligence is made up of simple reactive agents, which are able to perform only very simple functions (hash bruteforce, for example). Besides, swarm intelligence as a system, despite the fact that it is decentralized, is able to solve a very narrow range of tasks, because every its agent performs the same operation in one moment of time (the principle of redundant paralleling of the tasks).

In contrast, the distributed intelligence (in the BOINC example) is able to solve a very wide range of tasks and currently it is actively used in different areas, but at the same time, it is strictly centralized and limited by design pattern.

Decentralized intelligence should combine the two systems, compensating for their disadvantages and increasing their benefits.

Decentralized intelligence SONM will use BOINC parts for distribution and integration of the tasks, as well as Ethereum smart contracts and p2p technology for decentralization.

**Scheme of implementation of decentralized solution "miner hub"**

Consider the process how miners and hubs communicate with each other when they need to establish cooperation, i.e., the phase when the miner has not yet decided whether to participate in the hub and receive tasks from him or not (Fig. 2.1).

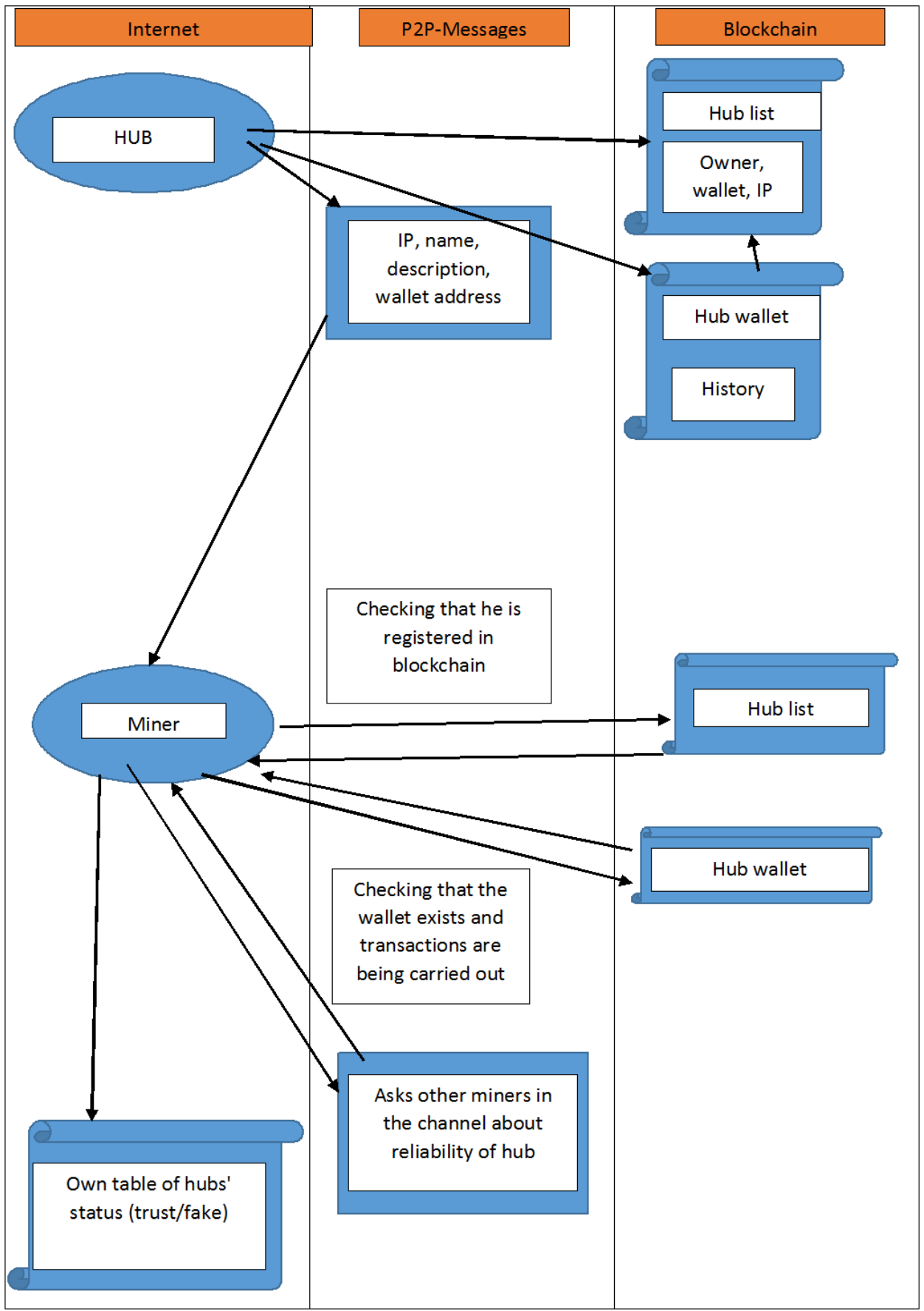

Fig. 2.1. Scheme of construction of the trust.

At the start of the hub administrator disposes in the blockchain network a smart wallet contract, where SONM tokens, which he will pay to miners, will be located. Then, address of this smart contract, as well as the address of the organizer and IP of hub are recorded in a special SONM smart contract "Hub Pool List", which includes unconfirmed (unverified) hubs in the form of events, and the proven, i.e., introduced to the white list (in the first version it will be prepared by SONM, and further it will be formed only by the miners) hubs are included as an array. In any case, information about the hub in SONM's smart contract will contain information about the address of the hub owner, address of hub wallet and IP address. In case of change of IP address, hub owner can change the record, as well as in case of changing of the wallet address.

That is, the hub registers the contract of the wallet, which contains the funds, which the hub pays to miners (so that the miners can check that the funds really exist) and registers basic information about itself, including the address of the owner and address of a wallet in a separate contract.

Then, the agent on the side of the hub starts broadcasting in the network of p2p messenger protocol, sending a message about itself in the format «IP, organizer address, wallet address, and name» into a common channel.

Agent on the side of the miner listens to the channel, and receives the message data from the servers, and then makes the request to the smart contract and compares data from the messages with data in the "white list". Miner may customize the agent so that he would accept messages from all servers or only from those, which are registered in smart contract as "proven".

After that, miner's blockchain agent asks the information about the contract-wallet of the hub, the number of money in the contract and a list of recent transactions of the wallet.

Intelligent analysis is conducted to compare the obtained data with the conditions, which were made by miner − whether funds in the hub wallet are sufficient and payments are regular, besides, the average amount of cash paid by hub is checked.

Then messenger agent sends a direct message to the hub to request additional meta-data, and records full information about the hub in its table with the mark "not confirmed".

At the same time, messenger agent regularly asks the questions to the common channel of miners with information about the hub, the average amount of remuneration paid to them, and so on, agents of other miners in the channel answer positively, if this information is correlated with their information, or negatively, if they believe this hub is not reliable.

In the case if the miner agent receives a sufficient amount of confirmations, it makes a mark against the hub in its table as "checked", and if the transaction, which has been received by a miner from this hub corresponds to the original agreements, then the status of a hub changes to "safe"

After that, depending on the settings of miner's software, the miner can either manually select a hub to which he can connect to perform tasks, or miner's agent will automatically select and connect to a hub, which offers the maximum profit.

A scheme of exchanging messages "miner-hub" is below (Fig. 2.2).

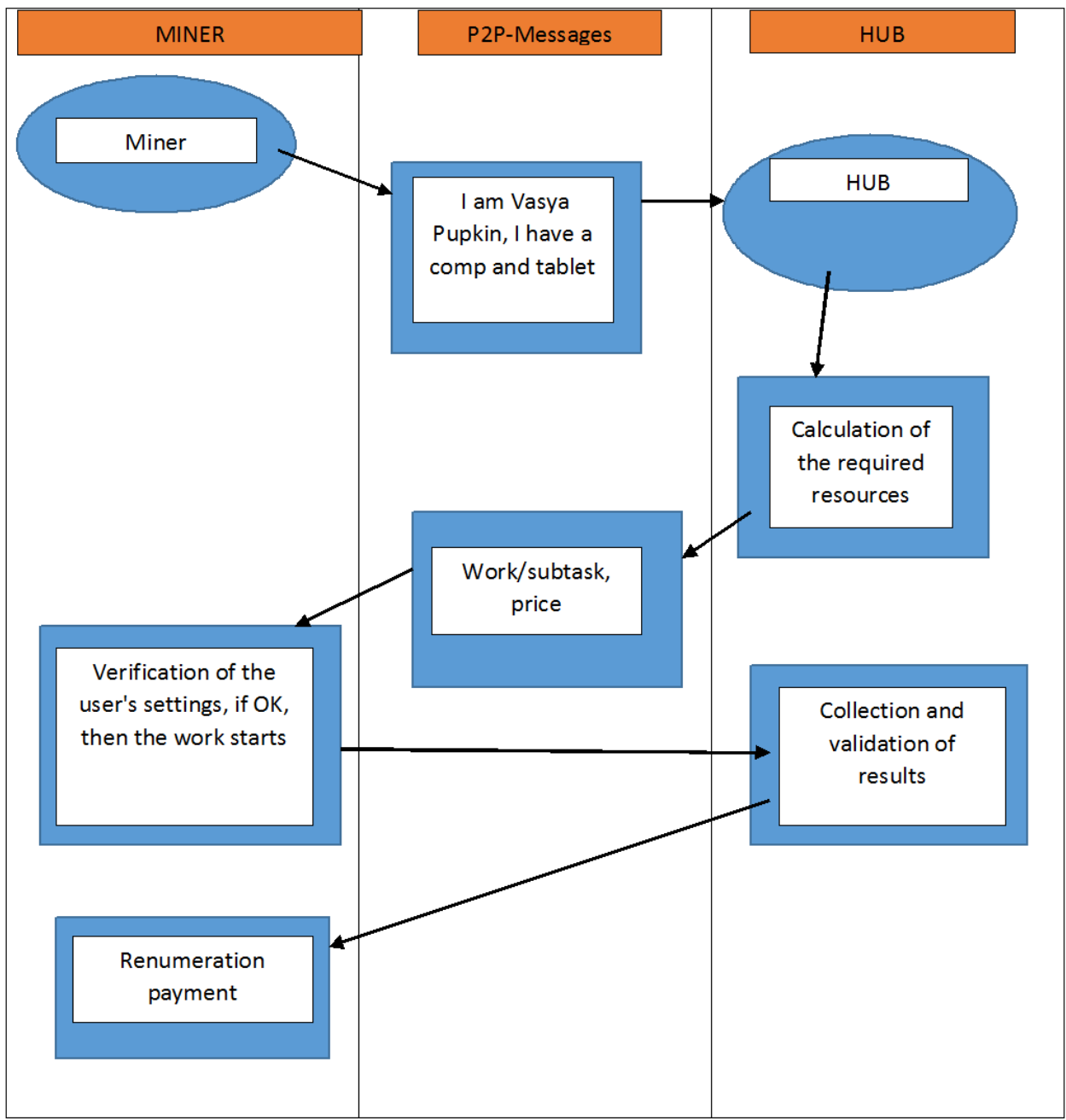

Fig. 2.2. Scheme of interrelations between miner and hub.

Task distribution itself is carried out, considering maximizing the efficiency of the equipment used, so the miner will always get the task, which is most suitable for him. The distribution of tasks and validation of the results is performed by standard means of BOINC.

Validation using a method of supernodes occurs in the channel, which is created by miners of one hub, working on the same task. Within the channel, the miners broadcast hashes of the results obtained, and simultaneously listen to messages from other miners. In the case if the result is the same, then they rebroadcast the message, including in it a confirmation of the result. In the case of reaching a consensus, the result is sent to the server with a mark of consensus, while in case of conflict detection, all versions of the results are sent to the server with a mark of conflict. In the case of conflict, the payment of remuneration for work is delayed, and a hub in the expert mode checks the result for compliance with the canonical approximation method. More information about the expert mode of the results testing can be found in BOINK documentation.

By default (in the normal operation mode without supernodes), hubs always work in expert mode.

**Scheme of the implementation of decentralized solution "client-hub"**

Work of client agents (the buyers) with the hubs is substantially similar to the miner agents, excluding the intellectual parsing of the results, which in this case prefers the hubs with the maximally lowest price (and vice versa in the case of miners). It is more probable that the clients will use "Application Pool" (stated in the white paper), than a contract "Hub Pool" (Fig. 2.3).

A method of content delivery is the only significant difference that we make in this process.

As you might expect, there no difference between the rendering of 6-hour video in the local computer and *uploading* of this video to the server with waiting of the end of video rendering in the server, because most part of the time will be used for uploading.

We have also the solution for this problem: when a client wants to upload a large file of raw data to the server, it simply creates a new torrent-file and sends a message indicating the operation to the selected hub. The hub receives the message and creates a task sequence for the uploading of this torrent, implementation of the work with the file and creation of a new torrent of the obtained file. The hub sends information about new file to the buyer, who only has to download the received file from the miners. We expect that this will be the most rapid solution of all existing ones at the moment.

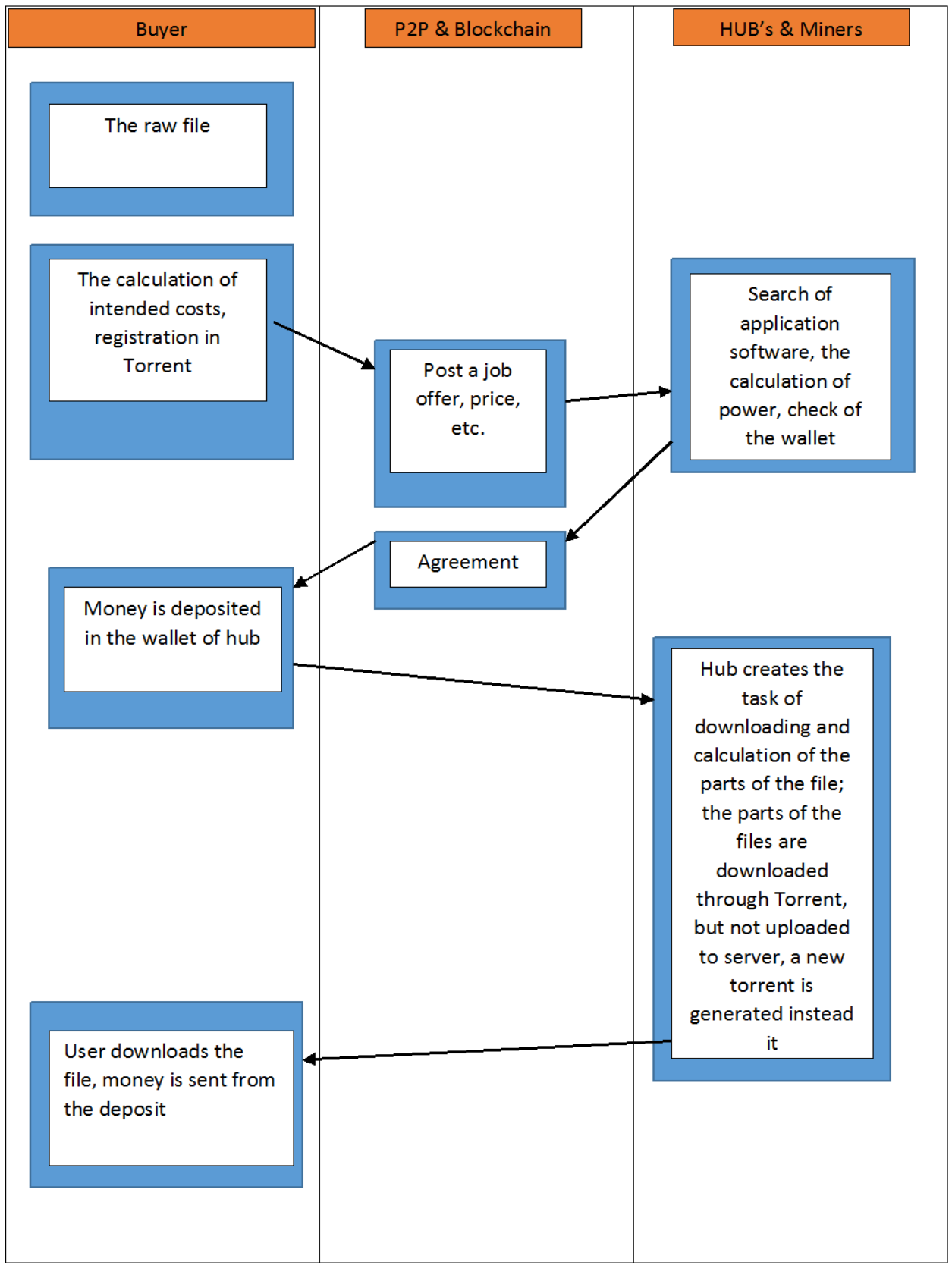

Fig. 2.3. Scheme of work with clients in the case of decentralized solution "client-hub". Some intermediate messages in the scheme are omitted.

Money is sent to the contract of hub wallet at the opening of the tasks where it is deposited. When the buyer receives the result, he confirms the transfer of money by means of smart contract function (similar to a contract of Multisignature Wallet).

Note that all transactions between the agents, blockchain, etc, go "under the hood", so the end users see only what they need. That is, the same buyer can simply select the desired application and run it, and the rest of the process − from the hub search to an implementation of the work by miners and the assembly of the result − may occur invisibly for him. You give the command "Make for me a part of this form!" and the system searches hubs that work with modeling applications and give them the task of calculating the parts with specific functions. Further, the hub distributes the task between miners, those obtains the results and the client receives a drawing. You say, "Give a forecast of dollar course for two months forward" and the system with enormous power calculates the probability of currency courses' movement and gives you a prediction. You say, "Give me a drawing of three-storey house in the Japanese style" and the system gives you a drawing of a Japanese three-storey house. You say, "Who will be killed by J. Martin next" and the system searches for an application for the calculation of the probabilities (or even neural network!), feeds it with all of the books "Game of Thrones" and you get a ready answer in a few minutes. You point − I punch!

In other words, such a system would be able to solve absolutely any problems with virtually unlimited power (and with the appropriate application on the hubs, of course) and maximum efficiency.